\PassOptionsToPackage{colorlinks, citecolor=cyan, linkcolor=purple, urlcolor= blue}{hyperref}
\documentclass[twocolumn, switch]{article} % Method A for two-column formatting

\usepackage{preprint}

\usepackage{amsmath, amsthm, amssymb, amsfonts}
\usepackage[version=4]{mhchem}

\usepackage[numbers,square]{natbib}
\usepackage{xcolor}		% colors for hyperlinks
\usepackage{booktabs} 		% professional-quality tables
\usepackage{nicefrac}		% compact symbols for 1/2, etc.
\usepackage{float}			% Allows for figures within multicol
\usepackage{lipsum}		%  Filler text

\usepackage{siunitx}
\usepackage{stfloats}

\usepackage{newfloat}
\DeclareFloatingEnvironment[name={Supplementary Figure}]{suppfigure}
\usepackage{sidecap}
\sidecaptionvpos{figure}{c}

\usepackage{titlesec}
\title{Refinement and Performance Benchmark for Range-Separated Water Force Field}

\usepackage{xwatermark}
\newwatermark[firstpage,color=gray!90,angle=0,scale=0.28, xpos=0in,ypos=-5in]{*correspondence: \texttt{jmchen32@gmail.com, kuangyu.2025@bytedance.com}}

\usepackage{authblk}

\author[1]{Qian Gao}
\author[1\thanks{\tt{jmchen32@gmail.com}}]{Junmin Chen}
\author[1,2\thanks{\tt{kuangyu.2025@bytedance.com}}]{Kuang Yu}

\affil[1]{Institute of Materials Research, Tsinghua Shenzhen International Graduate School (TSIGS), Shenzhen, PR China.}
\affil[2]{Bytedance Seed - AI for Science, Shenzhen, PR China.}
\usepackage{graphicx} % Required for inserting images
\usepackage{amsmath}
\usepackage{xr}
\usepackage{comment}

\usepackage{booktabs}
\usepackage{url}
\usepackage{adjustbox}
\usepackage{array}
\usepackage{xcolor}

\begin{document}

\twocolumn[ % Method A for two-column formatting
  \begin{@twocolumnfalse} % Method A for two-column formatting
  
\maketitle

\begin{abstract}

In our previous work, we developed a CCSD(T)-level range-separated water force field that combines the power of physics-driven and machine learning models. However, it was found that expensive CCSD(T)/CBS calculations lead to limited number of QM data as well as the missing of force labels, both of which lead to training instability issues. Bulk properties show large variations that cannot be resolved by simply reducing the fitting error in small cluster QM dataset. Such instability in bulk phase simulation is a universal problem in the training of machine learning potentials (MLPs), and is particularly severe at CCSD(T) level of theory.
%The range-separated water force field proposed by some of us has built a promising framework for constructing accurate and transferable molecular force field combining the power of physics-driven models in the long range and machine learning models in the short range. However, there lack systematic MD benchmarks for key macroscopic properties such as density. Besides, the prohibitive scaling of CCSD(T)/CBS method constrains the dataset can be generated for training the short-range ML model, which hinder the enhancement of stability and accuracy of PES.
In this work, using our range-separated water model as an example, we aim to overcome these limitations by developing a new training workflow. It is composed by several techniques including: 1. an active learning protocol that ensures more thorough sampling in different temperatures and densities; 2. an intermediate force label technique employing machine learning density functional; and 3. an ensemble knowledge distillation (EKD) method. These techniques significantly stabilize the resulting water model, consistently achieving sub-chemical accuracies in both cluster energies and experimental properties. Benchmarks are carried out for various properties including densities, radial distribution functions (RDFs), dielectric constants, diffusivity, and infrared spectra, all showing state-of-the-art (SOTA) performances and proving the effectiveness of the training protocol. 

%In particular, we retrain the short-range machine learning model with an augmented dataset, which encompasses chemical-accuracy machine learning surrogate forces integrated with CCSD(T) energies by transfer learning and more diverse data sampled across various temperatures and pressures by active learning protocols. Furthermore, we decrease the standard deviation of macroscopic properties predicted by the model ensemble with the ensemble knowledge distillation (EKD) method. Finally, we demonstrate that the refined water force field not only achieves sub-chemical accuracy in cluster energies, but also agrees with experiment in predictions for various properties of liquid water, such as density, RDFs, dielectric constant, diffusion coefficients, and infrared spectra. 
\end{abstract}
%\keywords{First keyword \and Second keyword \and More} % (optional)
\vspace{0.45cm}
  \end{@twocolumnfalse} % Method A for two-column formatting
]% Method A for two-column formatting

%\begin{multicols}{2} % Method B for two-column formatting (doesn't play well with line numbers), comment out if using method A

%%%%%%%%%%%%%%%  Main text   %%%%%%%%%%%%%%%
% \linenumbers

\section{Introduction}

Water is one of the most common substances in nature and, despite its seemingly simple molecular structure, it exhibits a wide range of anomalous behaviors\cite{brini2017water} arising from its complex hydrogen-bond network\cite{eisenberg2005structure}. Therefore, water has served both as a starting point and as a stringent benchmark for testing force field development strategies. The performance of the water force field is very sensitive to the accuracy of the underlying training data\cite{tian2024dynamic,chenDataEfficient2023}. Therefore, ideally, one should use the gold-standard CCSD(T)/CBS level training labels to faithfully capture the intricate many-body interactions in the water system. However, CCSD(T) labels are prohibitively expensive for large molecules and periodic systems due to the notorious O($N^7$) scaling of CCSD(T). This poses a great challenge, as one needs to learn from a limited dataset of small clusters in vacuum, and generalize it to bulk simulation, which is a highly nontrivial task.

Until now, many works have been done to develop CCSD(T) level water force field models. The many-body expansion (MBE) approach, such as MB-pol\cite{zhu2023mb,palos2024current} and q-AQUA-pol\cite{yu2022q,qu2023interfacing,yuStatus2023}, achieved remarkable agreement with experimental measurements of various properties. However, the complexity of PIP functions and the number of many-body terms increase rapidly with system size and chemical complexity, limiting its application in general multicomponent molecular systems. While machine learning potential (MLP) provides a promising alternative, it usually requires large amounts of energy/force data from periodic bulk-like structures, which is only affordable at the density functional theory (DFT) level. Some attempts have been made to improve the accuracy of MLPs from DFT to CCSD(T) level: for example, transfer learning can be performed using CCSD(T) labels of very few periodic systems of affordable size\cite{chenDataEfficient2023}. Alternatively, $\Delta$-learning can be performed targeting the difference between CCSD(T) and DFT labels for small finite clusters\cite{meszaros2025short,o2025towards}. In typical settings, pure MLPs without explicit long-range terms are local inherently and cannot correctly describe long-range interactions beyond their cutoff radius\cite{unke2021machine}, so efforts have been made to incorporate long-range interactions into the MLPs, such as DPLR\cite{zhang2022deep}, FFLUX\cite{symons2022application}, CACE-LR\cite{cheng2025latent,king2025machine}, etc.

Along this line, in our previous work\cite{yangTransferrable2022b}, we proposed a range-separated water force field that carefully partitions the intermolecular interactions based on distance. The long-range interactions are described using the physics-motivated components (i.e., electrostatic, polarization, and dispersion), parameterized by rigorous TD-DFT and SAPT calculations\cite{schmidt2015transferable,mcdaniel2016next}. The short-range interactions are described using conventional isotropic pairwise additive terms\cite{van2016beyond} in conjunction with an anisotropic many-body machine learning (ML) correction\cite{zhang2019embedded}, both parameterized using CCSD(T) data. The long-range part of the model adopts asymptotic parameters determined by the linear response properties of a single molecule, which is valid across various physical environments ranging from clusters in vacuum to condensed phase. Meanwhile, the ML model accounts only for short-range physics; thus, in principle it can be trained on small clusters where CCSD(T) labels are affordable. This range-separation method, later extended to the "PhyNEO" scheme, has shown excellent performance in large molecules including polymers\cite{chen2023phyneo,tu2025enhancing} and multicomponent systems such as electrolytes\cite{chen2025hybridphysicsdrivenneuralnetwork}, demonstrating its great capability. 

% Nevertheless, there are still unsolved problems for the previous water force field, particularly relevant to the training of short range machine learning model. 

However, our PhyNEO water force field was limited by poor training stability, which can be observed in more extensive bulk phase benchmarks. For sensitive bulk properties (e.g., densities, diffusivity, IR spectra), it was found that smaller errors in the QM training or validation data set do not always guaranty improvements in bulk property predictions\cite{fuForcesAreNot2023}. Consequently, performing comprehensive MD benchmarks for various properties of condensed phase is essential to validate the physical fidelity of the developed potential\cite{chen2025application}. This is especially an important issue for training at the CCSD(T)/CBS\cite{halkier1999basis} level, due to the small sample size and the lack of force labels. 
As for CCSD(T)/CBS calculations, analytical energy gradients (forces) are often prohibitively expensive or not implemented in standard quantum chemistry packages\cite{werner2012molpro,neese2020orca}. Meanwhile, previous studies show that force labels are extremely helpful in reducing the error in energy fitting \cite{christensen2020role}, so the absence of CCSD(T)/CBS force can be detrimental to the development of high-level MLPs. Some work combined lower-level force and higher-level energy to train MLPs with multi-fidelity learning and achieved high accuracy\cite{messerly2025multi}. However, more consistent force and energy labels are always beneficial for efficient training. 

%Nevertheless, overcoming these limitations is beneficial, as both force information and extensive data coverage are valuable for ensuring the accuracy and stability of the potential energy surface (PES). 
%The benefits can be understood intuitively. First, gradients provide the curvature information of PES that ensures the physical smoothness of PES. Besides, force vectors, unlike a single scalar energy value, provide $3N$ components for a $N$-atom configuration. A previous study suggests that one force label contributes as effectively to the convergence of prediction errors as $3N$ energy labels\cite{christensen2020role}.  % It has been reported that incorporating force labels can improve the accuracy in frequency, zero point energy\cite{} and vibrational spectroscopy calculation\cite{}. 
%Second, a large, comprehensive dataset ensures that the high-dimensional potential energy surface is densely populated, effectively transforming the learning task from risky extrapolation to safer interpolation\cite{zhang2020dp}. 

Recently, ML \textit{ab initio} methods such as DeePKS\cite{chen2020deepks}, % which is a delta-learning machine learning framework based on generalized kohn-sham functional and trained with CCSD(T) labels,
provide a feasible way to generate chemical-accuracy labels at the scaling of lower level DFT methods. Using a ML density functional, ML-DFT can learn from high-level data (e.g., CCSD(T)) and augment the data set with a comparable accuracy but with a much more favorable computational cost\cite{li2022deepks+}.  In this work, DeePKS is used as a surrogate model for CCSD(T) to generate cheap pretrain force data, while a transfer learning (TL) approach is used to ensure the CCSD(T) accuracy of the final model.

%Alternatively, we can also integrate the surrogate gradients with CCSD(T) energies via a Transfer Learning (TL) approach.
% Data amount is very important for improving the stability and accuracy of PES, 
Besides force data from ML-DFT, we also implemented two other data augmentation techniques to enhance the stability of the final model: active learning (AL)\cite{kulichenko2024data} and ensemble knowledge distillation (EKD)\cite{gong2025predictive}. In each cycle of AL, multiple models are trained independently using the currently available data. Trajectories are generated using the trained models and the deviations between different models are used to indicate new data points to be added into the training set. More samplings are also conducted in various temperatures and densities, so the training data achieve a better coverage over different physical conditions. Moreover, averaged predictions over all models are also generated, which are then learned by new models in an EKD procedure. We show that such an EKD approach greatly enhances the model stability in bulk property predictions. Overall, combining ML-DFT force label with AL and EKD methods, we eventually reach a much more robust training strategy. The performance and stability of the resulting hybrid water force field are demonstrated using not only cluster energies, but also various bulk properties.

We emphasize here that the problems addressed in this work are of general importance for high-level MLP development. Water is chosen as a showcase, because its anomalous bulk behaviors are well-known to be sensitive to the quality of PES. This is a typical challenging example where the CCSD(T) accuracy of PES is needed for quantitative predictions. However, high-quality data are sparse and expensive to obtain at the CCSD(T) level, which constitutes the central dilemma in modern MLP development. The techniques used in this work, including the range-separated architecture of the PhyNEO model, the intermediate force label using ML-DFT, and the AL and EKD techniques, all contribute to the solution of this central dilemma. Only facilitated by these techniques can we obtain a SOTA water model using only ~$3\times10^4$ CCSD(T) data in clusters up to only pentamers. Therefore, the training procedure developed in this work is important in all scenarios where the availability of high-level ab initio data is limited.

\section{Methods}
\label{sec:headings}

\subsection{Force Field Architecture}

% \lipsum[7] See Section \ref{sec:headings}.
% \paragraph{Paragraph}
% \begin{equation}
% \xi _{ij}(t)= {\frac {\alpha _{i}(t)a^{w_t}_{ij}\beta _{j}(t+1)b^{v_{t+1}}_{j}(y_{t+1})}{\sum _{i=1}^{N} \sum _{j=1}^{N} \alpha _{i}(t)a^{w_t}_{ij}\beta _{j}(t+1)b^{v_{t+1}}_{j}(y_{t+1})}}
% \end{equation}

Here, we briefly encapsulate the architecture of our PhyNEO water force field detailed in the previous paper\cite{yangTransferrable2022b}. The force field is composed of a single molecule intramolecular term and intermolecular interaction terms, as given by Eqn. \ref{eqn:Etot}. The intramolecular term corresponds to the spectroscopically accurate potential for a single \ce{H2O}\cite{partridge1997determination} molecule. While the intermolecular interaction can be decomposed into three parts, as given by Eqn. \ref{eqn:Einter}, the long-range ($E_{lr}$) and short-range isotropic pairwise ($E_{sr,iso}$) terms described by physically motivated functions, while the short-range anisotropic term ($E_{sr,aniso}$) is represented by the Embedded Atom Neural Network (EANN)\cite{zhang2019embedded} model. 

\begin{equation}\label{eqn:Etot}
E_{tot}=E_{intra}+E_{inter}
\end{equation}
\begin{equation}\label{eqn:Einter}
E_{inter}=E_{lr}+E_{sr, iso}+E_{sr, aniso}
\end{equation}

More specifically, long-range interactions ($E_{lr}$) are composed of electrostatic, polarization, and dispersion. In large distances, each term is determined by the asymptotic atomic parameters: atomic charges ($q_i$),  multipoles ($M_i$), polarizabilities ($\alpha_i$) and dispersion coefficients($C_{n,i}$). All these parameters can be rigorously obtained using linear response calculations (i.e., TD-DFT) in conjunction with population analysis (i.e., ISA-pol\cite{misquitta2016ab,misquitta2018isa}) and multipole expansion. These long-range terms are damped in short-range, in combination with the Born-Mayer charge penetration terms with parameters (e.g., exponents $B_i$ and prefactors $A_i$) fitted to CCSD(T) data\cite{hesselmann2005density,van2016beyond}. The function forms of all these terms are given by the following equations:
%The isotropic pairwise short-range terms employ the form of the Born-Mayer function, and parameters A are fitted to CCSD(T) data. Each term is described by equations (2-10).

\begin{equation}
E_{lr}=E_{es}+E_{pol}+E_{disp} \\
\end{equation}
\begin{equation}
E_{es}=\sum_{i<j} M_i^T T_{i j} M_j + \sum_{i<j} f_1\left(B_{i j}, r_{i j}\right) \frac{q_i q_j}{r_{i j}} \\
\end{equation}
\begin{equation}
B_{i j}=\sqrt{B_i B_j} \\
\end{equation}
\begin{equation}
f_n(\lambda, r)=1-e^{-\lambda r} \sum_{m=0}^n \frac{(\lambda r)^m}{m!} \\
\end{equation}
\begin{equation}
E_{pol}=\sum_{i \neq j} \mu_i^{i n d} T_{i j} M_j \\
\end{equation}
\begin{equation}
\mu_i^{\text {ind}}=\alpha_i\left(\sum_j T_{i j} M_j+\sum_j T_{i j} \mu_j^{i n d}\right) \\
\end{equation}
\begin{equation}
E_{disp}=-\sum_{i<j} \sum_{n=6,8,10} f_n\left(B_{i j}, r_{i j}\right) \frac{C_{n, i j}}{r^n} \\
\end{equation}
\begin{equation}
C_{n, i j}=\sqrt{C_{n, i} C_{n, j}} \\
\end{equation}
\begin{equation}
E_{sr, i s o}=\sum_{i<j} A_{i j} \exp \left(-B r_{i j}\right)
\end{equation}

\begin{figure}
  \centering
%   \subcaptionbox{Water hexamers\label{fig:subfig-a}}
    % {\includegraphics[width=\linewidth]{figures/workflow_new_f.pdf}}
    {\includegraphics[width=\linewidth]{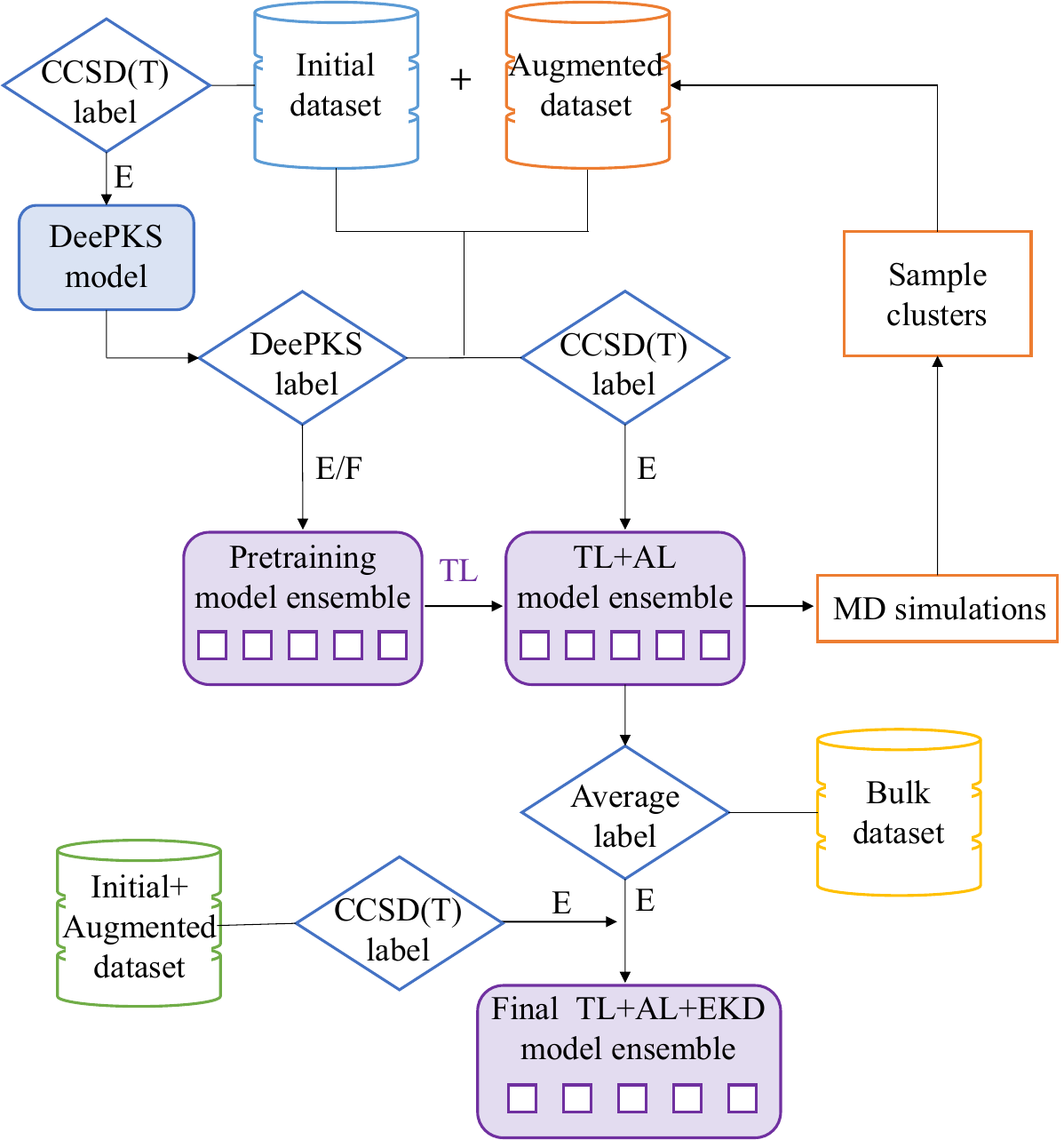}}
%   \subcaptionbox{Water dodecamers\label{fig:subfig-b}}
%     {\includegraphics{figures/fig12mer.png}}
  \caption{The workflow of the training protocol. First, a ML-DFT (DeePKS) model is trained using the initial CCSD(T) energy data and then is used to generate cheap energy/force (E/F) labels. The models are pretrained using DeePKS E/F labels and then transferred to CCSD(T) level using the original CCSD(T) E labels. Active learning using force prediction uncertainties are then conducted iteratively to generate an augmented dataset and the TL+AL model ensemble. In the final stage, the TL+AL model ensemble is distilled using bulk data to generate the final TL+AL+EKD model ensemble.}
  \label{fig:workflow}
\end{figure}

The physically motivated part of the force field, including both the pairwise isotropic short-range and the long-range components, is unchanged in this work. Here, we focus on the retraining of the ML correction part ($E_{sr,iso}$) modeled by EANN. An overview of the training workflow is depicted in Fig. \ref{fig:workflow} and we will give a detailed introduction in the next section.

\subsection{ML Model Training}

\subsubsection{DeePKS Labels}
CCSD(T) force calculations are expensive, so we use DeePKS as our surrogate model for force label generation. DeePKS is a general-purpose ML Kohn-Sham functional method, that can be trained using CCSD(T) data and gives predictions at comparable accuracy but with DFT-level computational cost\cite{chen2020deepks,chen2023deepks}. And since DeePKS is formally a DFT calculation, it is much more straightforward to compute force compared to naive CCSD(T).
%DeePKS is a generalized Kohn-Sham functional ML model that can predict with the precision of CCSD(T) and the computational cost of DFT\cite{chen2020deepks,chen2023deepks}. 
% We take the CCSD(T)-F12a/CBS data from our previous paper\cite{yangTransferrable2022b} as the initial dataset in this work, and use it to first train a DeePKS model. 
The training dataset for DeePKS model includes monomers, dimers, trimers, tetramers, and pentamers for training and validation, and hexamers, octamers, and dodecamers for testing. Training and validation configurations are randomly split with a ratio of 9:1. % which are sampled from qtip4p/f Path Integral Molecular Dynamics (PIMD) simulation with 32 beads at 300K 1bar 
% The original training/validation set consists of 1421 dimers, 2214 trimers, 2090 tetramers, and 10647 pentamers. 801 monomers are added to enhance the prediction accuracy of DeePKS in intramolecular energies, which will be used when computing intermolecular interaction energy. Furthermore, the intermolecular interaction calculations require the use of ghost atoms (i.e., off-center basis functions) to correct the basis set superposition error (BSSE)\cite{kestner1999basis}. To enhance the performance of DeePKS in the calculations with counterpoise corrections, we also randomly select 10\% of the data from the original training set, and randomly remove some molecules to create 1639 monomers with ghost atoms. These data with ghost atoms are critical for the stable performance of DeePKS in intermolecular energy calculations. 
%and 801 monomers and 1639 monomers with ghost atoms randomly selected from 10\% of the dimer to pentamer dataset. 

In detail, the original training/validation set consists of 1421 dimers, 2214 trimers, 2090 tetramers, and 10647 pentamers, which come from our previous work\cite{yangTransferrable2022b}. Then we added 801 monomer data with geometries sampled from a 300 K PIMD simulation of a single water molecule. Furthermore, intermolecular interaction calculations often require the use of ghost atoms (i.e., off-center basis functions) to correct the basis set superposition error (BSSE)\cite{kestner1999basis}. To enhance the performance of DeePKS in the calculations with counterpoise corrections, we also randomly select 10\% of the data from the original training set and randomly remove some molecules to create 1639 monomers with ghost atoms.

%Monomer with ghost atoms is a monomer from a larger cluster (for example, a dimer) with ghost atoms,i.e., an atom with zero nuclear charge and no electrons, placed at the positions of the other monomer's nuclei and carrying the other monomer's basis functions. It is a computational trick in counterpoise correction that is used to correct for the Basis Set Superposition Error  (BSSE)\cite{kestner1999basis}. % The inclusion of monomers with ghost atoms is for the purpose of learning BSSE, though we did not use the intermolecular interaction from DeePKS, but the total interaction due to the disagreement between analytical and numerical value of some force components of monomers with ghost atoms. 

The training of the DeePKS model was executed by iteratively calling the two procedures: solving the self-consistent field (SCF) equations and training a neural network energy functional. In the SCF calculation, we used aug-cc-pVDZ basis set. In iterative learning, the batch size was set at 16, and the total number of training epochs was set at 1500. The learning rate started at 0.0003 and decayed by a factor of 0.95 for every 300 epochs. After training for 20 iterations, we used the model in the 15th iteration as our final DeePKS model to generate training labels for the ML (EANN) part of the PhyNEO model. The choice was made to avoid overfitting since there was no significant improvement in the validation errors in subsequent iterations. 

We used the final DeePKS model to calculate the total energy and force for 1408 dimers, 2193 trimers, 2072 tetramers and 9112 pentamers selected from the DeePKS training set, in total, 14785 configurations. This dataset then served as the initial pretraining dataset for PhyNEO (\textit{vide infra}). The total energies of 500 octamers, 8 most stable hexamers and 7 most stable dodecamers were calculated to benchmark the testing performance of the DeePKS model. % These two tests are different, since octamers represents bulk-like structures, and hexamer and dodecamer represents cluster-like structures. 
The performance of the final DeePKS model is shown in the Supporting Information S\uppercase\expandafter{\romannumeral1}.

\subsubsection{Transfer Learning}

After we obtained the DeePKS model at the CCSD(T) level, we pretrained the short-range anisotropic EANN term of PhyNEO, using the DeePKS energy and force data as labels. The anisotropic intermolecular interaction energy labels were obtained by subtracting the $E_{lr}$ and $E_{sr,iso}$ terms calculated by DMFF\cite{wang2023dmff} and the monomer energies calculated by MBX\cite{riera2023mbx} from the total DeePKS energy, as shown in Eqn \ref{eqn:Etot}. Subsequently, a TL process was conducted using the original CCSD(T)-F12a/CBS intermolecular energy data. Such a TL process ensures that the final model is still at the CCSD(T) level and avoids the accumulation of errors from the DeePKS training process.
%Then we did transfer learning with short range anisotropic intermolecular interaction energy on the CCSD(T) level, that is, the CCSD(T)-F12a/CBS intermolecular interaction energy calculated by Molpro subtracted the long-range and short-range pairwise intermolecular interaction energy, as shown in equation (1), in order to enhance the model accuracy and alleviate the potentially existing BSSE. 

% \begin{equation}\label{eqn:Etot}
% E_{tot}=E_{intra}+E_{lr}+E_{sr, i s o}+E_{sr, a n i s o}
% \end{equation}

The input configurations of the training process are the DeePKS benchmark dataset mentioned above. All 14785 dimer, trimer, tetramer, and pentamer data points are used for training, while 500 octamer data points are used for validation. % Afterwards, we test the EANN model performances in the hexamers and dodecamers dataset.

% Firstly, we conduct the training of EANN models with the DeePKS energy and force labels. Forces are the negative gradient of potential energy and provide the curvature information of the energy landscape. We expect that training with both energy and force labels will benefit the fitting of PES by enforcing consistency between energy and forces. 

During the pretraining process, we set the energy and force prefactors in the loss function to be 0.1 and 0.05, respectively, to balance the weights of energy and force. The neural network architecture contains two hidden layers with 64 neurons for each input and hidden layer. The training was carried out with a maximum training epoch of 1200. We trained 5 parallel models, forming a model ensemble. Then, with each trained model in DeePKS level as a starting point, we conducted 5 parallel TL to CCSD(T)-F12a/CBS level to obtain 5 TL models. The hyperparameters were kept the same except for the prefactor of force was set to 0 as no force labels were available at the TL stage. These training processes usually stopped within 400 epochs.

% For the sake of comparison, we trained 5 parallel EANN models with the same CCSD(T)-F12a/CBS energy labels and hyperparameter settings starting from a randomly initialized neural network parameters. 

\subsubsection{Active Learning}

We further augmented the dataset with the AL strategy, which includes 3 iterative stages: training, exploration, and labeling. 

As stated in the previous section, we first trained 5 parallel models at DeePKS level and then transferred them to the CCSD(T) level with the initial 14785 data points.

In the exploration stage, employing a randomly selected model from the 5 trained models, we conducted classical MD simulations using 512 water molecules to generate new cluster configurations. The MD simulations were conducted using the NVT ensemble, under 15 different physical conditions covering 3 temperatures, i.e., 280 \si{K}, 300 \si{K} and 320 \si{K} and 5 densities, i.e., 0.90 \si{g/ml}, 0.96 \si{g/ml}, 1.00 \si{g/ml}, 1.04 \si{g/ml}, and 1.10 \si{g/ml}. % We select frames from the trajectories after equilibrium with a time interval of 2 ps. For MD conducted at densities of 0.96 \si{g/ml}, 1.00 \si{g/ml}, and 1.04 \si{g/ml}, we utilize the last 100 ps as production run, while for others we utilize the last 50 ps.
The local uncertainties of the force were then computed as shown in Eqn. \ref{eqn:local_uncertainty} and \ref{eqn:s_in}\cite{heid2024spatially}  for each oxygen atom $j$ in each selected frame $i$ of the MD trajectories.

%\begin{equation}\label{eqn:local_uncertainty}
%s_{ij}^{\text{local}} = \frac{1}{N_n} \sum_{n}^{N_n} \frac{1}{3} \sum_{x,y,z} s_{in}^{k}
%\end{equation}
\begin{equation}\label{eqn:local_uncertainty}
s_{ij}^{\text{local}} = \frac{1}{N_j} \sum_{n\in\mathcal{N}(j)}^{N_j} \frac{1}{3} \sum_{k\in\{x,y,z\}} s_{in}^{k}
\end{equation}
%\begin{equation}\label{eqn:s_in}
%s_{in}^{k} = \sqrt{ \frac{1}{N} \sum_{m}^{N} \left( F_{in}^{km} -  \overline F_{in}^{k} \right)^2 }
%\end{equation}
\begin{equation}\label{eqn:s_in}
s_{in}^{k} = \sqrt{ \frac{1}{N} \sum_{m}^{N} \left( F_{in}^{km} -  \overline F_{in}^{k} \right)^2 }
\end{equation}

Here, the local uncertainty around the particle $j$ ($s_{ij}^{local}$) is defined as the averaged atomic force uncertainties ($s_{in}^k$) over all its neighbors within  4 \si{\angstrom} (denoted as $n\in\mathcal{N}(j)$), and $N_j$ is the number of neighbors of the particle $j$. The average is also conducted on all three spatial dimensions, with index $k$ denoting the dimension (i.e., $x, y, z$). Meanwhile, the uncertainty of the atom force $s_{in}^k$ is defined as the standard deviation of the force predictions among all 5 parallel models in the model ensemble. In Eqn. \ref{eqn:s_in}, $m$ is the model index and $N=5$ is the number of parallel models. $F_{in}^{km}$ then denotes the $k$-component force of atom $n$ predicted by model $m$ in frame $i$, while $\overline F_{in}^{k}$ is its average over all models.

%Here, $s_{ij}^{local}$ denotes the local uncertainty for the $j$-th atom in the $i$-th data point. 
%It is obtained by averaging the uncertainty over $N_n$ neighboring atoms $n$ within 4 \si{\angstrom} around the atom $j$, and averaging over $k$ spatial direction $x$, $y$, $z$. The uncertainty of the force $s_{in}^{k}$ is the standard deviation of the atomic forces predicted by the model ensemble. $N$ is the number of parallel models, $\overline F_{in}^{k}$ is the average of the model ensemble predictions of the force.

As was previously demonstrated\cite{heid2024spatially}, the local uncertainty defined by the above equation is correlated with the actual local error. Therefore, we randomly selected oxygen centers with local uncertainties within the quantile range [0.950, 0.999]. Then we carved out a cluster of water around central oxygen, which includes all neighboring water molecules with an O-O distance smaller than 3.4 \si{\angstrom}. This approximately includes the first solvation shell in liquid water\cite{eisenberg2005structure}. On the basis of these clusters, we further randomly remove redundant neighboring water molecules to obtain a balanced sampling over dimer, trimer, tetramer, and pentamer configurations. 

%Afterwards, within the environment of 3.4 \si{\angstrom} cutoff radius around the selected oxygen centers, which is close to the first solvation shell of liquid water\cite{eisenberg2005structure}, we randomly sampled neighboring waters and obtained dimer, trimer, tetramer, and pentamer configurations. %with the ratio of 1:1.5:2:2.5.

In the labeling stage, we computed the energies and forces of the sampled water clusters using DeePKS and the energies using CCSD(T), so both the pretraining and the TL datasets are enhanced. 
% We calculated labels in DeePKS level after cleaning the not converged data points.

Finally, in the training stage, we repeated the aforementioned pretraining and TL processes at DeePKS/CCSD(T) levels to generate 5 new independent parallel models. And the exploration-labeling-training cycle was conducted for 6 iterations. Overall, 17626 new water cluster configurations were generated, including 2425 dimers, 4362 trimers, 4689 tetramers, and 6150 pentamers. Combining with the initial 14785 data points in the initial training set, it leads to a total set of 32411 data points. The analysis of the data distribution and stability tests are presented in the Supporting Information S\uppercase\expandafter{\romannumeral2}. 

%In total, we conducted active learning for 6 iterations and collected 17626 water cluster configurations as well as energy and force labels, including 2425 dimers, 4362 trimers, 4689 tetramers, and 6150 pentamers. Adding to the initial 14785 clusters dataset, there are 32411 water clusters altogether. The analysis for the data distribution and stability tests is presented in the Supporting Information S\uppercase\expandafter{\romannumeral2}. % For further details about the sampling condition constitution of water cluster candidates, please refer to the Supporting Information.

\subsubsection{Ensemble Knowledge Distillation}

In order to further reduce the variation of the MD simulation results given by the model ensemble, %trained by the augmented 32411 clusters dataset, 
we adopted the ensemble knowledge distillation (EKD) method proposed by Gong et al.\cite{gong2025predictive}. 
In this process, we performed 512 water simulations using 5 parallel models, in both NPT and NVT ensembles at temperatures of 277 \si{K}, 298 \si{K}, and 318 \si{K}, respectively, generating 30 simulations. For each simulation, we randomly sampled 10 frames from a 200 \si{ps} production run, generating 300 bulk frames.
%We sampled 10 frames from each of the MD simulation trajectories of 512 waters driven by the 5 parallel EANN models, conducted in the NPT and NVT ensemble and at temperatures of 277 \si{K}, 298 \si{K}, and 318 \si{K}, respectively, each with a production run of 200 \si{ps}. 
%Then we labeled the 300 bulk frames with the averaged energy predicted by the model ensemble. These 300 bulk data are then combined with existing CCSD(T) cluster data with an effective weighting factor of 32 for each bulk data point. 
These 300 bulk data are then combined with the existing CCSD(T) cluster data to form the new dataset. 
Then all 5 parallel models are finetuned using this combined dataset so they learned the distilled information from the model ensemble.

\subsection{MD Benchmarks}

MD simulations for density, dielectric constant, and RDF were carried out using the PhyNEO potential implemented in the DMFF software\cite{wang2023dmff} interfaced with i-PI 3.0\cite{litmanIPI2024}. For self-diffusion coefficients and infrared spectra, we utilized the  NVT/NVE integrator from our previous work\cite{han2025refining} instead of i-PI to propagate the dynamics. All MD simulations were performed with a time step of 0.5 \si{fs}. The MD simulations in the NPT ensemble were performed with Langevin thermostat and Bussi-Zykova-Parrinello barostat, and the MD simulations in the NVT ensemble were performed with Langevin thermostat.

To assess the MD stability of the model, we performed NPT MD simulations for all 5 parallel models in each model ensemble with 512 water molecules in a cubic box at 277 \si{K}, 298 \si{K}, and 318 \si{K} (at 1 atmosphere) and calculated the densities from the equilibrated trajectories for each model, respectively. The simulation time lengths and density values are listed in Table S5 and S6 in Supporting Information S\uppercase\expandafter{\romannumeral3}. The variation of density results are then used to quantify the stability of the training process. 

After verifying the stability of the training process, we chose the model with the best density performance for all following property benchmarks. The densities of liquid water were calculated using 512-water MD simulations at a series of temperatures including 248 K, 268 K, 277 K, 298 K, 318 K, 338 K, and at 1 atmosphere pressure. For each temperature, the first 300 ps of the generated trajectory was considered as the equilibration process, and the rest of the trajectory was used for production analysis. Standard uncertainties were calculated using the block averaging method. Details of simulation time lengths and density values are listed in Table S7 in Supporting Information S\uppercase\expandafter{\romannumeral4}.

% based on a 95$\%$ confidence interval
For the calculation of RDFs, we performed MD simulation at 298 K in the NVT ensemble for 200 ps and computed the RDFs with MDAnalysis\cite{gowers2019mdanalysis} from the last 100 ps trajectory after equilibration. The simulated system consists of 512 water molecules in a cubic box, with the density set to the experimental density of 0.997 \si{g/ml}.

For the calculation of the dielectric constant, MD simulations were performed in NVT at 298 K with 256 water molecules for 7.5 \si{ns} and the last 6.5 \si{ns} of the trajectory were used for the production analysis. The initial structures were extracted from the last 20 ps of a 200 ps NPT MD trajectory, for which the densities were within the standard deviation of the average density predicted by the model. Three independent simulations were performed to obtain the average value and standard deviation.

For dynamic properties, i.e., diffusion coefficients and infrared spectra, we first ran a 500 ps MD simulations with 256 molecules in the NVT ensemble at 298 K and the experimental density in 1 atm. Subsequently, we ran 30 MD NVE simulations, starting from the initial configurations uniformly drawn from the last 300 ps NVT trajectory with an interval of 10 ps. Each NVE MD simulation was performed for 20 ps and the velocities and positions were saved every 1 step. Three independent experiments were conducted to obtain the average value and standard deviation.
% The uncertainty of the self-diffusion coefficient was calculated by bootstrapping recommended by reference\cite{maginn2019best}.

\section{Results and Discussion}
\label{sec:others}

% \subsection{The Inconsistency Problem and Solution}

\subsection{Model Ensemble Performance on Water Clusters}
%Energies and Forces}

To show the stability of the training process, we compare four model ensembles: the ones trained with the initial CCSD(T)/CBS dataset are denoted as "Initial", the ones trained with only TL and AL techniques are denoted as "TL+AL", the ones trained using the initial CCSD(T)/CBS dataset and EKD technique are labeled as "EKD", and the ones trained with both TL+AL and EKD (i.e., the full training scheme) are labeled as "TL+AL+EKD" hereafter. 

\begin{figure*}[h]
  \centering
%   \subcaptionbox{Water hexamers\label{fig:subfig-a}}
    {\includegraphics[width=\linewidth]{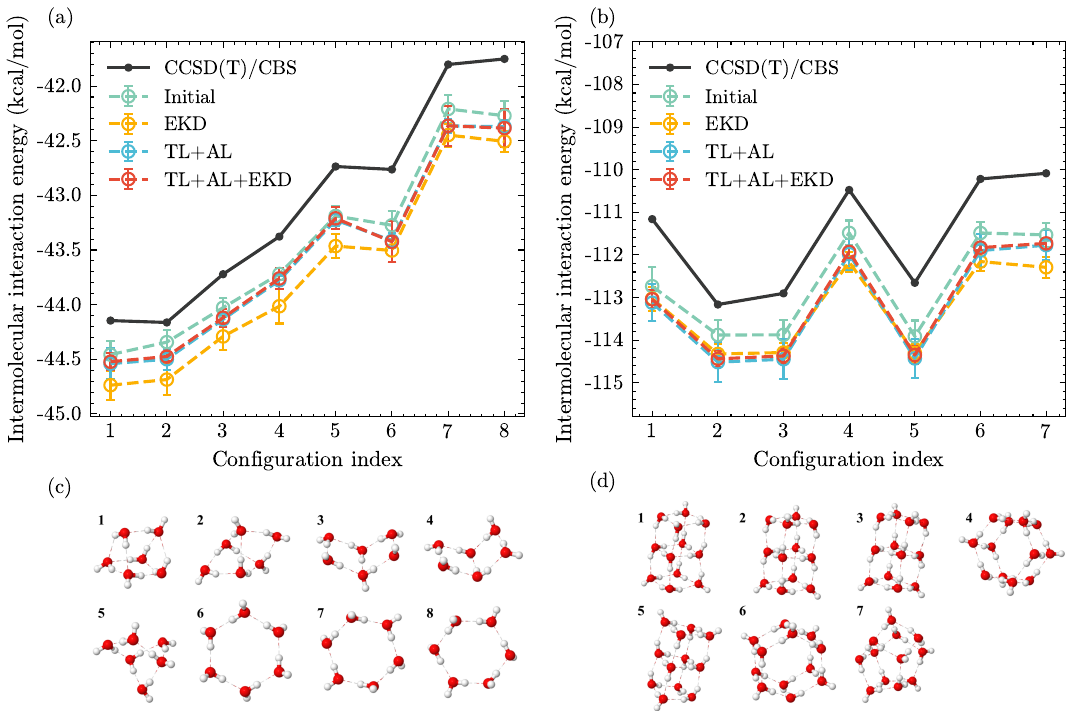}}
%   \subcaptionbox{Water dodecamers\label{fig:subfig-b}}
%     {\includegraphics{figures/fig12mer.png}}
  \caption{The intermolecular interaction energies predicted by the Initial, EKD, TL+AL, and TL+AL+EKD PhyNEO models on (a) hexamers test set and (b) dodecamers test set, compared with CCSD(T)/CBS reference. (c) The molecular structures of 8 isomers of water hexamer. (d) The molecular structures of 7 isomers of water dodecamer.}
  \label{fig:cluster}
\end{figure*}

We first calculated the average training and validation energy RMSE for the Initial, EKD, TL+AL, and TL+AL+EKD PhyNEO model ensembles, and list the results in Table \ref{tab:rmse-EF}. We can observe that the energy errors given by all parallel models are near sub-chemical accuracy. Comparison with DeePKS force labels is also listed in the table (further information can be found in Supporting Information Table S\uppercase\expandafter{\romannumeral1}). 
%Although we do not have available CCSD(T)-12a force labels, we still compared the EANN predicted atomic forces with the DeePKS labels for which the accuracy was evaluated in Supporting Information Table S\uppercase\expandafter{\romannumeral1}.
It can be seen that the force RMSEs are around 1 \si{kcal \cdot mol^{-1}/\angstrom}. Compared to the Initial model ensemble, the energy and force RMSEs of the TL+AL model ensemble are slightly improved for both training and validation datasets. This can be attributed to the augmented cluster training data contributed by the TL+AL process. Otherwise, it can be seen that the EKD method does not significantly improve the energy and force RMSEs of the EKD and TL+AL+EKD model ensemble, compared to their base model ensembles (Initial and TL+AL), respectively, but rather slightly worsens the results. Nevertheless, all the training and validation errors of all model ensembles are rather similar. 

%However, as we will show below, models with very similar energy and force RMSEs can generate very different bulk simulation results, demonstrating the instability issue in ab initio MLP development.
%which incorporated bulk waters in the training process, slightly increases the training and validation RMSE. 
% energy of cluster or bulk water? DeePKS force is not reliable. 

\begin{table}[h]
  \centering
  \caption{The average RMSE of energy (\si{kcal \cdot mol^{-1}/atom}) and atomic force (\si{kcal \cdot mol^{-1}/\angstrom}) predictions of PhyNEO model ensembles for the training (denoted as trn) and octamer validation (denoted as val) set of water clusters. The $\mathrm{RMSE}^\mathrm{E}$ and $\mathrm{RMSE}^\mathrm{F}$ were calculated w.r.t. CCSD(T)/CBS labels and DeePKS labels, respectively.}
  \resizebox{0.5\textwidth}{!}{
  \begin{tabular}{lllll}
    \toprule
    Model Ensemble & $\mathrm{RMSE}^\mathrm{E}_\mathrm{trn}$ & $\mathrm{RMSE}^\mathrm{F}_\mathrm{trn}$ & $\mathrm{RMSE}^\mathrm{E}_\mathrm{val}$ & $\mathrm{RMSE}^\mathrm{F}_\mathrm{val}$      \\
    % dataset & RMSE(trn,E) & RMSE(trn,F) & RMSE(val,E) & RMSE(val,F)     \\ 
    \midrule
    Initial & 0.013\(\pm\)0.001 & 1.11\(\pm\)0.01 & 0.019\(\pm\)0.001 & 1.25\(\pm\)0.03 \\
    EKD & 0.016\(\pm\)0.001 & 1.14\(\pm\)0.04 & 0.021\(\pm\)0.001 & 1.29\(\pm\)0.03 \\
    TL+AL & 0.011\(\pm\)0.001 & 0.91\(\pm\)0.02 & 0.016\(\pm\)0.001 & 1.07\(\pm\)0.02 \\
    TL+AL+EKD & 0.012\(\pm\)0.001 & 0.93\(\pm\)0.04 & 0.017\(\pm\)0.001 & 1.09\(\pm\)0.05 \\
    \bottomrule
  \end{tabular}
  }
  \label{tab:rmse-EF}
\end{table}

% \begin{figure}
%   \centering
%   \subcaptionbox{Training set\label{fig:subfig-a}}
%     {\includegraphics{figures/figtrain_e.png}}
%   \subcaptionbox{Validation set\label{fig:subfig-b}}
%     {\includegraphics{figures/figtest_8mer_e.png}}
%   \caption{Intermolecular interaction energy performance of EANN model on training and validation set}
%   \label{fig:multi-image}
% \end{figure}

In addition to the training set containing only dimers to pentamers and the validation set containing only octamers, we also examined the model performance on the most stable geometries of hexamers and dodecamers. Unlike the octamer validation set sampled from MD, this test set represents the model performance around the energy minimum. The correct prediction to the relative stabilities of these gas phase clusters also reflects the model's capability in size extrapolation.

As seen in Fig. \ref{fig:cluster}, all model ensembles systematically underestimate the intermolecular interaction of hexamers and dodecamers.  In comparison, the Initial model ensemble performs the best in this test set, while the TL+AL model ensemble performs slightly worse than the Initial model ensemble, and the further inclusion of EKD (i.e., TL+AL+EKD) made negligible differences. The increased error caused by TL+AL can be attributed to the increased weights of clusters extracted from bulk MD in the AL-augmented training set. A large portion of clusters in the original dataset were sampled from cluster PIMD simulations in vacuum, with structures much more similar to the low energy hexamers and dodecamers in the test set. Meanwhile, bulk MD geometries are less compact, as indicated by the larger O-O-O angle distributions \cite{yangTransferrable2022b}. Therefore, an increased portion of bulk MD geometries slightly worsens the model performance in compact water clusters in vacuum. More interestingly, directly applying EKD on the Initial model ensemble without TL+AL gives a worse performance. Nevertheless, the RMSEs of all the model ensembles are within 0.06 \si{kcal/mol} per atom, all showing sub-chemical accuracy. From the results of octamer/hexamer/dodecamer tests, it is unclear which model performs the best, but as we will show in the following sections, these model ensembles lead to significant differences in bulk simulation, thus highlighting the stability issue.

\subsection{Density Variations among Model Ensemble}

\begin{figure}
  \centering
  \includegraphics[width=\linewidth]{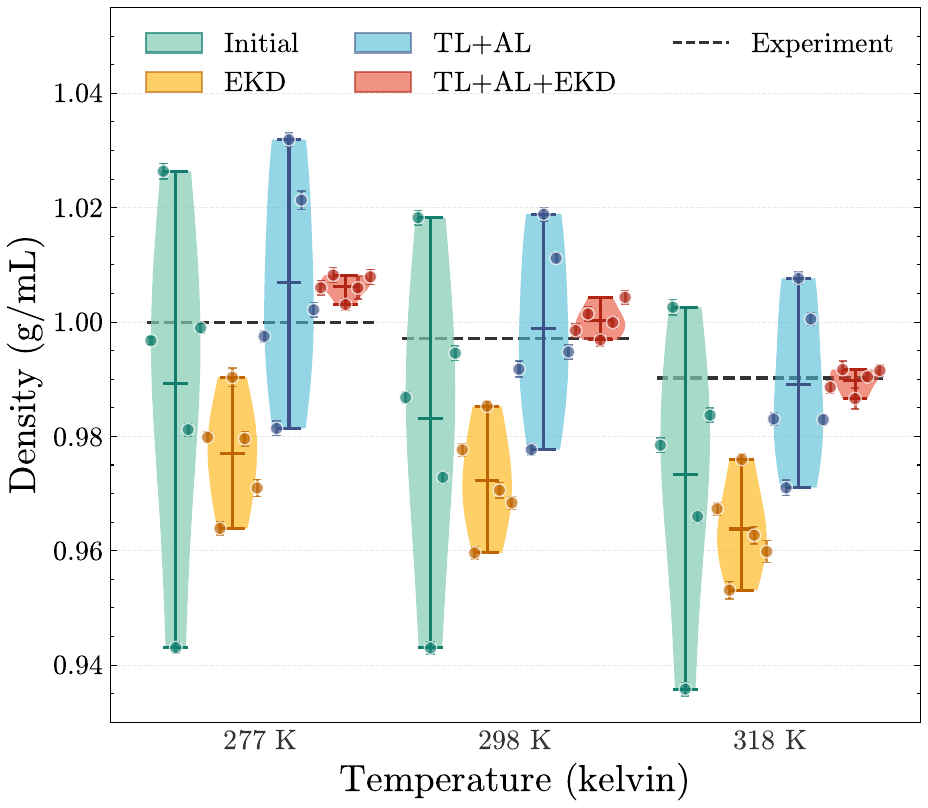}
  \caption{Comparison of the density distributions given by the Initial, EKD, TL+AL, and TL+AL+EKD model ensembles. The three groups of points and filled area from left to right correspond to simulated densities and the variations at 277 K, 298 K, and 318 K, respectively, compared with the experimental reference\cite{gs1975density} (black dashed lines).}
  \label{fig:density}
\end{figure}
% boxes with whiskers 

% We compared the densities at 277 K, 298 K, and 318 K predicted by the 5 parallel transfer learning models trained with the initial dataset and active learning dataset (TL+AL) with that predicted by the corresponding models finetuned by ensemble knowledge distillation (Distilled). 

In Fig. \ref{fig:density}, we compared the densities at 277 K, 298 K, and 318 K predicted by the 5 parallel models in the Initial, EKD, TL+AL and TL+AL+EKD model ensembles. 

First of all, we can see that the Initial model ensemble shows up to a variation of $\pm 4\%$ in the density prediction task, which is highly unstable. The EKD ensemble features a significant reduction in model variation to approximately $\pm 1\%$, but suffers from a severe systematic underestimation of $2-3\%$. The key role of EKD can be understood as the finite QM cluster dataset does not cover all the configurations encountered in bulk MD simulation. Nontypical structures (such as structures with shortened or elongated bonds or within repulsive intermolecular regions) are hard to sample in classical MD. They may not dominate the overall distribution, but they are still thermally accessible in MD, and the large errors in these structures lead to stability issues. As shown in Fig. \ref{fig:density}, EKD does not improve the averaged accuracy since no new physical data is introduced, but it indeed suppresses the random noises in the unsampled structures, thus improves the stability of the final model. Meanwhile, by further exploring the phase space and providing more actual E/F labels, the TL+AL model ensemble largely corrects the systematic error. However, it is still haunted by the stability issue, as indicated by the $\pm 3\%$ model variations shown in Fig. \ref{fig:density}. 
In comparison, the full training scheme (i.e., TL+AL+EKD) achieves excellent performance in both accuracy and stability simultaneously. The variation of the models within the ensemble is suppressed to less than $\pm 0.5\%$, with average errors less than 1\%. These results show the strong power of the combined TL+AL+EKD strategy.

%We can observe that both the Initial and the AL+TL parallel models exhibit significantly larger variations (up to $\pm 4\%$ and $\pm 3\%$, respectively) in the simulated densities. However, after the inclusion of EKD methods, the EKD parallel models show a much smaller variation of approximately $\pm 1\%$, and moreover, the TL+AL+EKD parallel models only vary within the range less than $\pm 0.5\%$. It is thus proved that the EKD method significantly improves the model stability in bulk property predictions. Otherwise, it can be seen that the average density of the EKD model ensemble, which involves EKD alone, is severely underestimated (by 2--3\%) compared to the experiment. While the TL+AL+EKD model ensemble behaves much better, showing density errors consistently within 1\%. Thus, we can assume that TL+AL are beneficial for the model accuracy in bulk property predictions.

Meanwhile, given the similar E/F RMSEs shown in the last section, it is clear that the fitting/validation/testing RMSEs in QM clusters are not good indicators for the quality of MLP, even though they are widely used in the current literature. Lower RMSEs in a limited QM dataset do not guaranty better performance in bulk simulations. Similar observations have been reported in a previous study\cite{fuForcesAreNot2023}, which investigated a number of existing MLPs and found that E/F RMSEs and bulk simulation metrics are not aligned. This inconsistency emphasizes the fact that careful benchmarks against high-quality experimental data are still indispensable.

%However, the obviously better density predictions made by the TL+AL+EKD model ensemble compared to that by the EKD model ensemble indicate that, actually, the TL+AL protocol should have sampled important physical regions in the PES. Additionally, the results imply that the EKD method requires a good baseline model ensemble to finetune, since, as seen in Fig. \ref{fig:density}, the TL+AL model ensemble shows both smaller density variations and average densities closer to experiments than the Initial model ensemble. As for the EKD model ensemble, the severe underestimation in density should probably due to the contribution of certain underperforming parallel Initial models to the averaged prediction labels.
%The probable explanation for this problem is the random error of ML models in the extrapolation region. Although the short-range EANN model encodes the local atomic environment within a 4 \si{\angstrom} cutoff, the finite water clusters are inherently different from periodic bulk waters. Consequently, applying the EANN models trained with water clusters to MD simulations of bulk waters introduced an extrapolation problem. The random errors in the extrapolation region propagate during the dynamics and lead to large standard deviations in macroscopic ensemble averages like density. Conversely, the ensemble knowledge distillation method, by finetuning the EANN models with the average energy prediction of the parallel AL+TL models for bulk waters, decreased the standard deviation by minimizing the random errors\cite{gong2025predictive}.

Nevertheless, both good accuracy and stability can be achieved when TL+AL is combined with EKD, validating the training procedure proposed in this work. Considering the computational cost, hereafter, we will use the best performer in the TL+AL+EKD model ensemble for all subsequent benchmarks.

\subsection{Bulk Phase Benchmarks for the Final Model}

In this section, we select the best performer of the TL+AL+EKD model ensemble to carry on with our benchmarks. We calculate various thermodynamic properties (density, radial distribution function, and static dielectric constant) and dynamic properties (self-diffusion coefficient and infrared spectrum) of liquid water and compare them with experiments. The benchmark results are also compared with other SOTA water potentials to show the quality of the training procedure. 

\subsubsection{Density}

\begin{figure}[h]
  \centering
  \includegraphics[width=\linewidth]{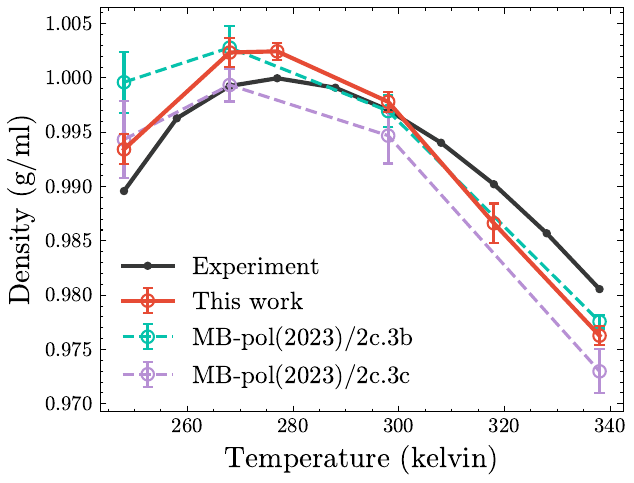}
%   \caption*{国外的期刊习惯将图表的标题和说明文字写成一段，需要改写为标题只含图表的名称，其他说明文字以注释方式写在图表下方，或者写在正文中。}
  \caption{Densities of liquid water at 248 K, 268 K, 277 K, 298 K, 318 K, 338 K, and at 1 atmosphere predicted by the final PhyNEO model in this work, compared with experiments\cite{gs1975density} and the reported results of MB-pol(2023)\cite{zhu2023mb}.}
  \label{fig:rho-T}
\end{figure}

Fig. \ref{fig:rho-T} shows the density as a function of temperature, in comparison to the experimental data and different versions of the MB-pol(2023) models. It can be seen that the densities agree well with experiments within an error of 0.005 \si{g/cm^{3}} throughout the simulated temperature range. The densities are close to the experiment at ambient temperature (298 K), whereas slightly overestimated at lower temperatures and slightly underestimated at higher temperatures. The overall accuracy is comparable with the SOTA MB-pol models, but with much less training data ($3\times 10^4$ CCSD(T) points versus $12\times 10^4$ CCSD(T) points). It is also encouraging to see that the density turning point at approximately 277 K is accurately captured, which is a well-known challenging task.

%Our simulations were conducted with classical MD simulations that neglect Nuclear Quantum Effects (NQE)\cite{ceriotti2016nuclear}, while previous works\cite{reddy2016accuracy,qu2023interfacing,o2025towards} suggested a decrease in density upon inclusion of NQE from path-integral molecular dynamics (PIMD) simulations. 

\subsubsection{Radial Distribution Function}

\begin{figure*} %[htbp]
  \centering
%   \subcaptionbox{O-O RDFs\label{fig:subfig-a}}
    {\includegraphics[width=\linewidth]{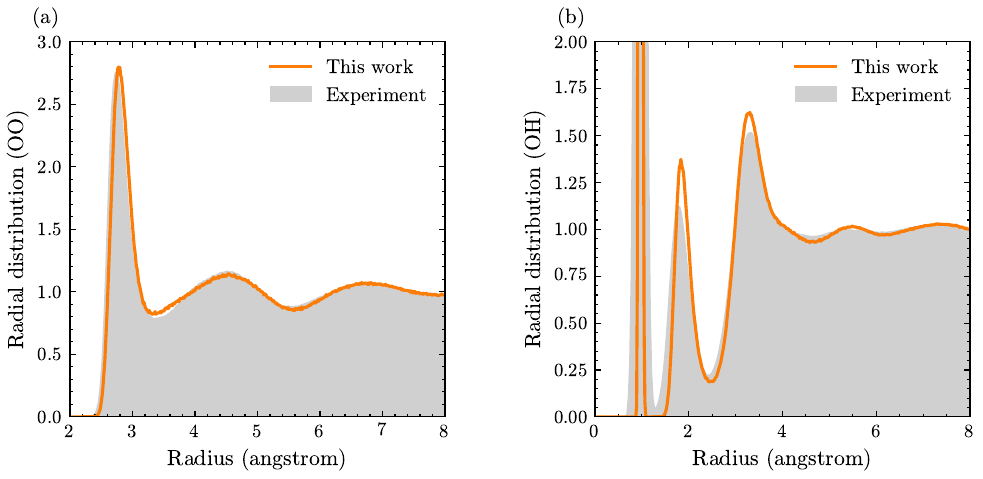}}
%   \subcaptionbox{O-H RDFs\label{fig:subfig-b}}
    % {\includegraphics{figures/output1.png}}
  \caption{OO radial distribution function (panel a) and OH radial distribution function (panel b) from MD simulations at 298 K using the PhyNEO model in this work, compared with experimental data\cite{soperRadial2000}.}
  \label{fig:rdf}
\end{figure*}
% \FloatBarrier

The OO radial distribution function (RDF) shown in Fig. \ref{fig:rdf}a is generally in good agreement with the experimental measurements.
% , though first peak () is slightly sharper and shifts towards longer OO distance and second peak () is more flat.
% we can observe that the amplitude of both the first peak in the calculated O-O RDF and the peaks in O-H RDF are higher compared to the experimental measurements. 
Meanwhile, the OH RDF in Fig. \ref{fig:rdf}b shows a more localized structure, with sharper peaks compared to the experimental measurements. This is most likely due to the negligence of nuclear quantum effects (NQE), as previous studies\cite{morrone2008nuclear,yangTransferrable2022b,yuStatus2023} indicated. Overall, the solvation structure of liquid water is well-reproduced.
% covalent peak (first peak), the hydrogen bonding peak (second peak) 
% This indicates that the prediction of water structures is more localized than that of real water.
%This should be alleviated by including NQE, as previous studies\cite{morrone2008nuclear,yangTransferrable2022b,yuStatus2023} reported that RDFs calculated from PIMD simulations that include the effects of NQE will lead to more delocalized RDFs.

% We can also observe that the postion of the first peak in O-O RDF and that of the second peak in O-H RDF shift slightly towards longer O-O, or O-H distance, which correspond to the first H-bonded neighbor. 

%\cite{skinner2013benchmark}

\subsubsection{Static Dielectric Constant}

The static dielectric constant, $\varepsilon$, is calculated from the fluctuations of the total dipole moment of the system:
\begin{equation}\label{eqn:eps}
\varepsilon=1+\frac{4\pi}{3 V k_B T}\left(\left\langle M^2\right\rangle-\langle M\rangle^2\right)
\end{equation}

where M is the total dipole moment of the simulation box and the angular brackets denote an ensemble average,  V is the volume of the simulation box, $k_B$ is the Boltzmann constant and T is the temperature. 

The focus of this work is to develop a high-accuracy potential energy surface (PES). But to accurately calculate the dielectric properties of water, we also need a reliable dipole moment surface (DMS). It is well-known that the optimal electrostatic/polarization model for PES model cannot be used directly to predict DMS. Therefore, we employ the ML Deep Dipole model from previous work \cite{zhang2020deep} to calculate the dipole moment in this work. %This model is based on deep neural networks (DNNs) and maximally localized Wannier functions (MLWFs).

\begin{table}[h]
  \centering
  \caption{Static dielectric constant of liquid water at 298 K from MD simulations with PhyNEO model in this work, compared with experiment\cite{archer1990dielectric} and MB-pol\cite{reddy2016accuracy}.}
  \begin{tabular}{lll}
    \toprule
    Potential          & Classical MD      & Experiment                   \\
    \midrule
    PhyNEO model   & 87 \(\pm\) 1   & 78.5 \\
    MB-pol(2016)   & 68.4                   \\
    \bottomrule
  \end{tabular}
  \label{tab:eps}
\end{table}

The dielectric constant calculated from the MD simulations of our model (Table \ref{tab:eps}) is 11$\%$ higher than the experimental value of 78.5, while the value reported by MB-pol is 13$\%$ lower than the experimental value. Therefore, the current PhyNEO model has an accuracy comparable to that of MB-pol (with a marginal improvement). NQE may affect the simulation results, but computation of the water dielectric constants requires tens of nanoseconds of MD sampling to converge\cite{raabe2011molecular,cai2025simulations}, due to the slow dynamics of the hydrogen bond network. Such a long path-integral quantum MD simulation is beyond our capability. Meanwhile, experimental dielectric constants for $\ce{H2O}$ and $\ce{D2O}$ are numerically close\cite{srinivasan1974pressure} (78.45 versus 78.08), indicating a minor NQE on water dielectric constants. 

%This should also be partly related to the neglect of NQE in our simulations. However, it has been observed that the dielectric constant typically requires a simulation time of tens of nanoseconds to converge\cite{raabe2011molecular,cai2025simulations}, thus the expense of PIMD simulations should be unaffordable, and few studies have reported the dielectric constant calculated from PIMD simulations.

\subsubsection{Self-diffusion Coefficient}

The self-diffusion coefficient of liquid water is calculated from the velocity autocorrelation function (VACF) as:
\begin{equation}
D=\frac{1}{3} \int_0^{\infty}\left\langle v(t) v(0)\right\rangle d t
\end{equation}

where $v$ is the velocity of the center of mass of the water molecule, and the angular bracket denotes the ensemble average. 

\begin{table}[h]
  \centering
  \caption{Self-diffusion coefficient (\si{\angstrom^2/\pico\second}) of liquid water at 298 K from MD simulations with the PhyNEO model developed in this work, compared with experiment\cite{holz2000temperature}, MB-pol\cite{reddy2016accuracy}, and q-AQUA-pol\cite{yuStatus2023}.} %(Å^2/ps)
  \begin{tabular}{lll}
    \toprule
    Potential          & Classical MD      &  Experiment                   \\
    \midrule
    PhyNEO model   & 0.183 \(\pm\) 0.003   & 0.230 \\
    MB-pol(2016)   & 0.23 \(\pm\) 0.02       \\
    q-AQUA-pol     & 0.185 \(\pm\) 0.004                     \\
    \bottomrule
  \end{tabular}
  \label{tab:diffusivity}
\end{table}

The calculated self-diffusion coefficient (see Table \ref{tab:diffusivity}) is underestimated compared to the corresponding experimental values. Compared to other models, PhyNEO performs worse than MB-pol, which gives an excellent description to water diffusion. Meanwhile, another SOTA CCSD(T) level water potential, q-AQUA-pol, also underestimates the self-diffusion coefficient in classical MD. And the results between PhyNEO and q-AQUA-pol are consistent. Interestingly, a significant increase was found in the self-diffusion coefficient of water after the inclusion of NQE through the TRPMD method\cite{qu2023interfacing,yuStatus2023}, due to the competing effects of intramolecular and intermolecular NQEs\cite{habershon2009competing}. Therefore, the underestimation of diffusivity is likely to be related to the negligence of NQE. Nevertheless, the accuracy of PhyNEO in water self-diffusion prediction is comparable with other SOTA water models at CCSD(T) level.

% This is in consistency with the over-structured feature of RDFs calculated by MD trajectory simulated with our force field model.

% For our model, Path Integral MD (PIMD) simulations are needed for further validation. 

\subsubsection{Infrared Spectrum}

The IR spectrum is calculated from the Fourier transform of the time autocorrelation function of the system’s dipole moments: %within the electric dipole approximation and linear response theory
\begin{equation}
\alpha(\omega) n(\omega)=\frac{2 \pi}{3 c \kappa_B T V} \int_{-\infty}^{\infty} d t e^{-i \omega t}\langle\dot{\mathbf{M}}(0) \cdot \dot{\mathbf{M}}(t)\rangle
\end{equation}

where $\alpha(\omega)$ is the absorption coefficient, $n(\omega)$ is the frequency-dependent refractive index, $c$ is the speed of light in vacuum, $\dot{{M}}(0) \cdot \dot{{M}}(t)$ is the classical equilibrium time correlation function (TCF) of the total dipole of the system, and the angular brackets denote an ensemble average over 30 independent MD trajectories. %with $\dot{{M}}(t)$ given by eq 2

\begin{figure}
  \centering
  \includegraphics[width=\linewidth]{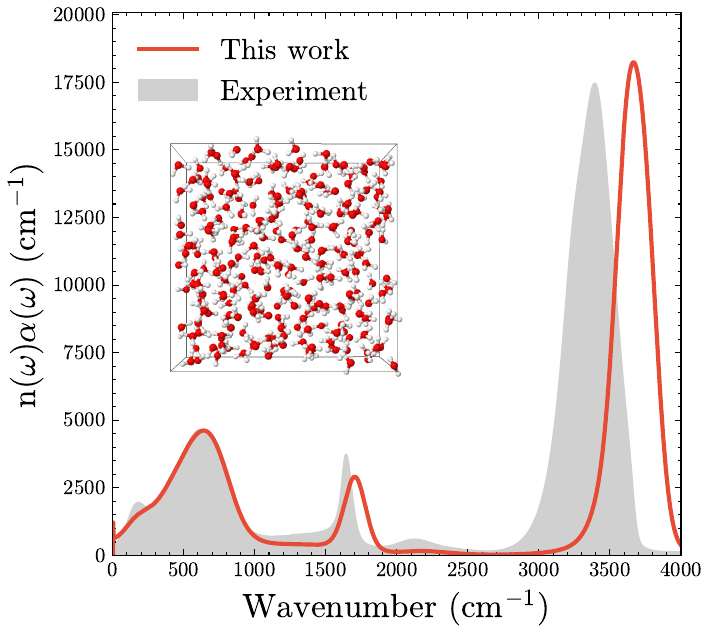}
  \caption{Infrared spectrum from MD simulations with PhyNEO model in this work, compared with experimental measurements\cite{bertie1996infrared}.}
  \label{fig:ir}
\end{figure}

There are 3 major bands in the experimental infrared spectra of liquid water, including the O–H stretching absorption band at around 3000-3800 \si{cm^{-1}}, a peak associated with H-O-H bending vibrations at around 1650 \si{cm^{-1}}, and intermolecular librational features below 1000 \si{cm^{-1}}. Fig. \ref{fig:ir} shows that the calculated spectrum reproduced the general features of the experimental spectrum\cite{bertie1996infrared}. It was demonstrated that the accuracy of both PES and DMS influences the lineshapes of the IR spectra, while PES determines the widths and positions of all spectra features, and DMS is crucial to capture the correct infrared intensity\cite{medders2015interplay}. We can observe that the H-O-H bending and O-H stretching bands of the PhyNEO spectrum are blue-shifted by ~50 and ~250 \si{cm^{-1}}, respectively, compared to their experimental counterparts. Similar observations were reported in reference\cite{reddyTemperaturedependent2017} by MB-pol. This is due to the neglect of NQE in classical dynamics simulations, which can be improved with quantum dynamics simulations. Meanwhile, the intermolecular librational band is in quite good agreement with the experimental measurements, showing the excellent accuracy of PhyNEO in intermolecular interactions. It is also noted that there is a small shoulder peak at around 200 \si{cm^{-1}}, which can be attributed to the charge transfer effects induced by the molecular vibration of water dimer along the OH hydrogen bond direction\cite{han2023incorporating}. Although the PhyNEO model shows a small shoulder at 200 \si{cm^{-1}}, the peak intensity is largely underestimated. Given that the DMS we use in this work can reproduce this shoulder peak in other MLPs\cite{zhang2020deep}, we tend to attribute this problem to the reduced vibrational magnitude in the corresponding mode in PhyNEO simulation. One possible reason is the PES inaccuracies in the tightly-bounded configurations with short intermolecular distance and possibly elongated OH covalent bonds. Considering that the initial dataset was generated using the empirical q-tip4p/f potential\cite{yangTransferrable2022b}, this region can be severely under-sampled. Corresponding improvement will be left to future work.

\section{Conclusions}

In summary, in this work we established a systematic training workflow that combines TL from intermediate ML-DFT force labels, AL, and ensemble knowledge distillation. This workflow aims to address the data deficiency and instability problem in the CCSD(T)-level MLP training. At this level, the force label is missing and the coverage of the dataset is often insufficient, making the bulk simulation results unstable even when small energy/force RMSEs in QM dataset are achieved. Using the range-separated PhyNEO water model, which is trained using limited small cluster (less than pentamer) data, we demonstrate the instability issue and justify the effectiveness of the new training protocol. The final ensemble of models reaches both good accuracy and small variation in both cluster energy and bulk density predictions. We further conducted a more comprehensive bulk benchmarks on the resulting model, showing that the final model achieves SOTA performance in various properties including density, RDF, dielectric constant, diffusivity, and IR spectrum. Meanwhile, the model is trained using much less data compared to other SOTA models, proving the efficiency of the current model and training scheme. The workflow established in this work can be beneficial to the development of more general high-level MLP models.

\section*{Supplementary Material}
The supplementary material provides the performance benchmarks for the DeePKS model, the distribution of the AL-augmented dataset, and the numerical values of the simulated densities in the results. The CCSD(T)/CBS energy data of all dimers, trimers, tetramers, pentamers, and octamers used for the training/validation of the EANN model are available at \url{https://github.com/plumbum082/water_force_field}. % It is a private repository now.
Other data that were generated and used in this study are available upon request to the authors. 
The final PES model and the code for the training workflow and MD benchmarks are available also at \url{https://github.com/plumbum082/water_force_field}.

%%%%%%%%%%%% Supplementary Methods %%%%%%%%%%%%
%\footnotesize
%\section*{Methods}

%%%%%%%%%%%%% Acknowledgements %%%%%%%%%%%%%
%\footnotesize
\section*{Acknowledgments}
We thank the National Natural Science Foundation of China (22473068) for the financial support of this work.

%%%%%%%%%%%%%%   Bibliography   %%%%%%%%%%%%%%
\normalsize
\bibliography{references}

%%%%%%%%%%%%  Supplementary Figures  %%%%%%%%%%%%
%\clearpage

%%%%%%%%%%%%%%%%   End   %%%%%%%%%%%%%%%%
%\end{multicols}  % Method B for two-column formatting (doesn't play well with line numbers), comment out if using method A
\end{document}

% \pagebreak
\widetext
% \onecolumngrid

\vspace{2ex}

\begin{center}
{\LARGE \textbf{Supplementary Information}}\\[3.5ex]
{\Large Refinement and Performance Benchmark for Range-Separated Water Force Field }\\[2.5ex]

\textbf{Qian Gao}$^{1}$,
\textbf{Junmin Chen}$^{1, *}$,
\textbf{Kuang Yu}$^{2, *}$ \\[1ex]
$^{1}$Institute of Materials Research, Tsinghua Shenzhen International Graduate School (TSIGS), Shenzhen, PR China.\\
$^{2}$Bytedance Seed - AI for Science, Shenzhen, PR China.\\
\textbf{E-mail: }\href{jmchen32@gmail.com}{jmchen32@gmail.com}, \href{mailto:kuangyu.2025@bytedance.com}{kuangyu.2025@bytedance.com}\\

% \usepackage{authblk}
% % \renewcommand*{\Authfont}{\bfseries}

% % % \author[1\thanks{\tt{asmith@college.edu}}]{Alice Smith\orcidA{}}
% % % \author[2]{Bob Jones\orcidB{}}

% \author[1]{Qian Gao}
% \author[1]{Junmin Chen}
% \author[1\thanks{\tt{yu.kuang@sz.tsinghua.edu.cn}}]{Kuang Yu}

% \affil[1]{Institute of Materials Research, Tsinghua Shenzhen International Graduate School (TSIGS), Shenzhen, PR China.}

\end{center}

% \vspace{2ex}

\newpage

%%%%%%%%%% Merge with supplemental materials %%%%%%%%%%
%%%%%%%%%% Prefix a "S" to all equations, figures, tables and reset the counter %%%%%%%%%%
\setcounter{equation}{0}
\setcounter{figure}{0}
\setcounter{table}{0}
\setcounter{page}{1}
\setcounter{section}{0}

\makeatletter
\renewcommand{\theequation}{S\arabic{equation}}
\renewcommand{\thefigure}{S\arabic{figure}}
\renewcommand{\thetable}{S\arabic{table}}
\renewcommand{\bibnumfmt}[1]{[S#1]}
\renewcommand{\citenumfont}[1]{S#1}
\renewcommand{\thesection}{S\Roman{section}}
%%%%%%%%%% Prefix a "S" to all equations, figures, tables and reset the counter %%%%%%%%%%

\section{DeePKS Model Performance}

\begin{figure}[htbp]
  \centering
  \includegraphics[width=0.8\linewidth]{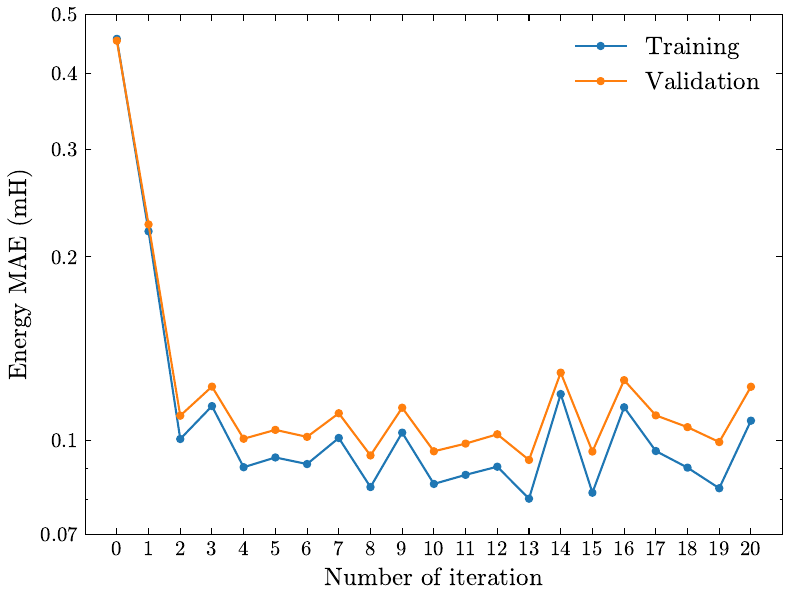}
  \caption{Energy MAE of DeePKS model during iterative learning.}%(The training iterations were executed with the unit Hartree.)
  \label{fig:trainlog}
\end{figure}

The mean absolute error (MAE) of the DeePKS model during iterative learning is shown in Fig. \ref{fig:trainlog}. We used the model in the 15th iteration as our final DeePKS model and benchmarked the prediction performance of total energy and force.

For energy predictions, the root mean squared error (RMSE) per atom for the selected 14785 training configurations is 0.006 \si{kcal/mol}, and that for the octamer test set is 0.013 \si{kcal/mol}, which is approximately two orders of magnitude lower than “chemical accuracy,” namely 1 \si{kcal/mol}. 

% shown in Figure S2,  and octamers testing because hexamers and dodecamers are the most stable geometries in 0K, while the clusters in training set and octamers in testing set are sampled from MD trajectories in 300 \si{K}. Their geometries have distinct characteristics, as the stable hexamers and dodecamers form more hydrogen bonds and shows lower density tetrahedral cage-like structures, vice versa, clusters sampled from 300 \si{K} do not form so many hydrogen bonds. %and shows higher density van der Waals contact structures. 
% Therefore, the increase in error should be acceptable and is still at least one order of magnitude lower than chemical accuracy and demonstrate a good transferability of DeePKS model.

% \begin{figure}
%   \centering
% %   \subcaptionbox{Water hexamers\label{fig:subfig-a}}
%     {\includegraphics[width=0.45\linewidth]{SIfigures/6tot_p.png}}
% %   \subcaptionbox{Water dodecamers\label{fig:subfig-b}}
%     {\includegraphics[width=0.45\linewidth]{SIfigures/12tot_p.png}}
%   \caption{Compare total energies calculated by DeePKS and CCSD(T)/CBS on test set}
%   \label{fig:multi-image}
% \end{figure}

\begin{table}[H]
  \setlength{\abovecaptionskip}{0.2cm}
  \setlength{\belowcaptionskip}{0.3cm}
  \setlength{\tabcolsep}{8pt}
  \centering
  \caption{The RMSE (\si{kcal \cdot mol^{-1}/\angstrom}) of DeePKS forces for several randomly selected water clusters, compared to numerical CCSD(T)-F12(a) forces. The subscript number for dimer, trimer, tetramer, and pentamer denotes the configuration index.}
  \begin{tabular}{lllllll}
    % \toprule
    \hline
    $\mathrm{dimer}_{118}$ & $\mathrm{trimer}_{710}$ & $\mathrm{tetramer}_{1656}$ & $\mathrm{tetramer}_{2071}$ & $\mathrm{pentamer}_{175}$ & $\mathrm{pentamer}_{6073}$ \\
    % \midrule
    \hline
    0.616 & 0.793 & 0.985 & 0.507 & 1.081 & 0.737  \\
    % \bottomrule
    \hline
  \end{tabular}
  \label{tab:forcermse}
\end{table}

For force predictions, due to the nonavailability of analytical CCSD(T)-F12(a) force for Molpro, we compared the DeePKS atomic forces with the numerical CCSD(T)-F12(a) forces calculated by Molpro. And we only benchmarked for several cluster configurations randomly selected from the training set, considering the high expense of CCSD(T) gradient calculations. It can be seen from Table \ref{tab:forcermse} that the force error is approximately around 1 \si{kcal \cdot mol^{-1}/\angstrom}.

\begin{figure}[H]
  \centering
  \includegraphics[width=\linewidth]{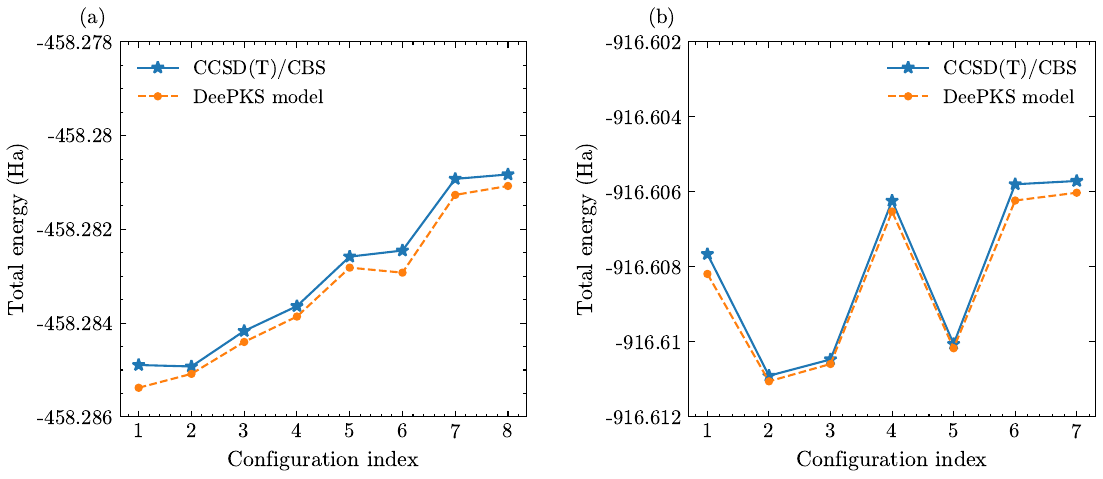}
  \caption{Compare total energies calculated by DeePKS model and CCSD(T)/CBS method on the hexamer and dodecamer test sets.}% (We plotted with the unit Hartree in to fit the magnitude of the numerical numbers.)
  \label{fig:testset}
\end{figure}

Fig. \ref{fig:testset} shows that the energies of the hexamer and dodecamer test sets are systematically underestimated by the DeePKS model. % This is not surprising since the hexamers and dodecamers are stable configurations, the structures of which are distinct from the octamers sampled from bulk MD trajectories.
The RMSEs per atom for hexamers and dodecamers are 0.076 \si{kcal/mol} and 0.052 \si{kcal/mol}, respectively. The energy RMSE of the DeePKS model is at least one order of magnitude lower than chemical accuracy for stable cluster test sets and demonstrates good transferability.

\newpage

\section{Active Learning Data Distribution}

\begin{table}[h]
  \setlength{\abovecaptionskip}{0.2cm}
  \setlength{\belowcaptionskip}{0.2cm}
  \setlength{\tabcolsep}{12pt}
  \centering
  \caption{Count of clusters with different number of monomers.}
  \begin{tabular}{lllll}
    % \toprule
    \hline
    dimer & trimer & tetramer & pentamer \\
    % \midrule
    \hline
    2425 & 4362 & 4689 & 6150 \\
    % \bottomrule
    \hline
  \end{tabular}
  \label{tab:xmer}
\end{table}

\begin{table}[h]
  \setlength{\abovecaptionskip}{0.2cm}
  \setlength{\belowcaptionskip}{0.2cm}
  \setlength{\tabcolsep}{12pt}
  \centering
  \caption{Count of clusters sampled from MD simulation trajectories of different densities (g/ml).}
  \begin{tabular}{llllll}
    % \toprule
    \hline
    0.90 & 0.96 & 1.00 & 1.04 & 1.10 \\
    % \midrule
    \hline
    2178 & 4382 & 4423 & 4433 & 2210 \\
    % \bottomrule
    \hline
  \end{tabular}
  \label{tab:Ps}
\end{table}

\begin{table}[h]
  \setlength{\abovecaptionskip}{0.2cm}
  \setlength{\belowcaptionskip}{0.2cm}
  \setlength{\tabcolsep}{12pt}
  \centering
  \caption{Count of clusters sampled from MD simulation trajectories at different temperatures.}
  \begin{tabular}{*{3}{c}} %{lllll}
    % \toprule
    \hline
    280 \si{K} & 300 \si{K} & 320 \si{K} \\
    % \midrule
    \hline
    5875 & 5885 & 5866 \\
    % \bottomrule
    \hline
  \end{tabular}
  \label{tab:Ts}
\end{table}

Table \ref{tab:xmer}, \ref{tab:Ps}, and \ref{tab:Ts} show the constitution of sampling conditions at which water cluster candidates are sampled during active learning (AL) protocols.

\begin{figure}[H]
  \centering
  \includegraphics[width=0.9\linewidth]{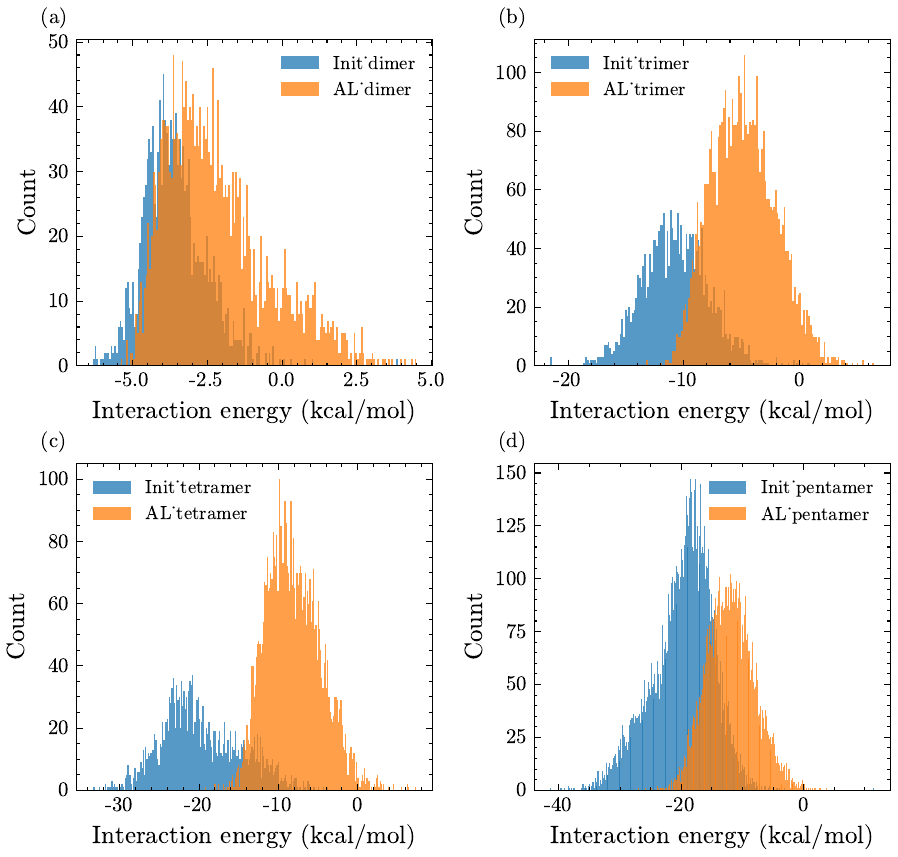}
  \caption{The CCSD(T) intermolecular interaction energy distributions in the initial training dataset (denoted as Init) and the AL-augmented dataset (denoted as AL).}
  \label{fig:distribution}
\end{figure}

% In order to gain a better understanding from the microscopic point of view, 
Additionally, we compared the distribution of CCSD(T) intermolecular interaction energy in the initial dataset (14785 configurations) and that in the AL dataset (17626 configurations). As seen in Fig. \ref{fig:distribution}, it indicates that we have sampled more clusters with higher intermolecular energy with AL compared to the initial dataset.

% \section{MD Stability Tests}
% \section{Effects of Active Learning on Stability of MD Simulations}

\section{Simulated Densities by Parallel Models}
% \section{Simulated Densities by Parallel Models before and after Ensemble Knowledge Distillation }

% \begin{table}[H]
%   \setlength{\abovecaptionskip}{0cm}
%   \setlength{\belowcaptionskip}{0.2cm}
%   \setlength{\tabcolsep}{3pt}
%   \centering
%   \caption{The simulated densities from MD simulations with five parallel models before (TL+AL) and after ensemble knowledge distillation (Distillation) and simulation time at 277 K, 298 K, and 318 K.}
%   \begin{tabular}{lllllllllllll}
%     % \toprule
%     \hline
%     \multirow{2}{*}{T (K)}  & \multirow{2}{*}{Time (ns)} & \multicolumn{5}{l}{TL+AL density (g/ml)}  & \multicolumn{5}{l}{Distilled density (g/ml)}  & \multirow{2}{*}{Exp. (g/ml)} \\
%     \cline{3-12}
%     & & m0 & m1 & m2 & m3 & m4 & m0 & m1 & m2 & m3 & m4 & \\
%     % \midrule
%     \hline
%     277	& 1.5 &	0.997 &	0.981 &	1.032 &	1.021 &	1.002 &	1.008 &	1.007 &	1.002 &	1.006 &	1.009 &	  1.000 \\ 
%     298	& 1.5 &	0.992 &	0.978 &	1.019 &	1.011 &	0.995 &	1.000 &	1.003 &	0.998 &	1.001 &	1.002 &	  0.997 \\
%     318	& 1.0 &	0.983 &	0.971 &	1.008 &	1.001 &	0.983 &	0.989 &	0.992 &	0.987 &	0.990 &	0.992 &	  0.990 \\
%     % \bottomrule
%     \hline
%   \end{tabular}
%   \label{tab:three-line}
% \end{table}

\begin{table}[H]
  \setlength{\abovecaptionskip}{0.2cm}
  \setlength{\belowcaptionskip}{0.2cm}
  \setlength{\tabcolsep}{3pt}
  \centering
  \caption{The calculated densities from MD simulations with five parallel models in the Initial and the EKD model ensemble, respectively, compared with experiments\cite{gs1975density}, and the simulation time lengths at 277 K, 298 K, and 318 K.}
  \begin{tabular}{lllllllllllll}
    % \toprule
    \hline
    \multirow{2}{*}{T (K)}  & \multirow{2}{*}{Time (ns)} & \multicolumn{5}{l}{Initial density (g/ml)}  & \multicolumn{5}{l}{EKD density (g/ml)}  & \multirow{2}{*}{Exp. (g/ml)} \\
    \cline{3-12}
    & & m0 & m1 & m2 & m3 & m4 & m0 & m1 & m2 & m3 & m4 & \\
    % \midrule
    \hline
    277	& 1.5 &	0.997 &	1.026 &	0.943 &	0.981 &	0.999 &	0.980 &	0.964 &	0.990 &	0.980 &	0.971 &	  1.000 \\ 
    298	& 1.5 &	0.987 &	1.018 &	0.943 &	0.973 &	0.995 &	0.978 &	0.960 &	0.985 &	0.971 &	0.968 &	  0.997 \\
    318	& 1.0 &	0.978 &	1.003 &	0.936 &	0.966 &	0.984 &	0.967 &	0.953 &	0.976 &	0.963 &	0.960 &	  0.990 \\
    % \bottomrule
    \hline
  \end{tabular}
  \label{tab:three-line}
\end{table}

\begin{table}[H]
  \setlength{\abovecaptionskip}{0.2cm}
  \setlength{\belowcaptionskip}{0.2cm}
  \setlength{\tabcolsep}{3pt}
  \centering
  \caption{The calculated densities from MD simulations with five parallel models in the TL+AL and the TL+AL+EKD model ensemble, respectively, compared with experiments\cite{gs1975density}, and simulation time lengths at 277 K, 298 K, and 318 K.}
  \begin{tabular}{lllllllllllll}
    % \toprule
    \hline
    \multirow{2}{*}{T (K)}  & \multirow{2}{*}{Time (ns)} & \multicolumn{5}{l}{TL+AL density (g/ml)}  & \multicolumn{5}{l}{TL+AL+EKD density (g/ml)}  & \multirow{2}{*}{Exp. (g/ml)} \\
    \cline{3-12}
    & & m0 & m1 & m2 & m3 & m4 & m0 & m1 & m2 & m3 & m4 & \\
    % \midrule
    \hline
    277	& 1.5 &	0.997 &	0.981 &	1.032 &	1.021 &	1.002 &	1.006 &	1.008 &	1.003 &	1.006 &	1.008 &	  1.000 \\ 
    298	& 1.5 &	0.992 &	0.978 &	1.019 &	1.011 &	0.995 &	0.999 &	1.001 &	0.997 &	1.000 &	1.004 &	  0.997 \\
    318	& 1.0 &	0.983 &	0.971 &	1.008 &	1.001 &	0.983 &	0.989 &	0.992 &	0.987 &	0.990 &	0.992 &	  0.990 \\
    % \bottomrule
    \hline
  \end{tabular}
  \label{tab:three-line}
\end{table}

\section{Final Model Performance on Density}
\begin{table}[H]
  \setlength{\abovecaptionskip}{0.2cm}
  \setlength{\belowcaptionskip}{0.2cm}
  \setlength{\tabcolsep}{8pt}
  \centering
  \caption{The calculated densities from MD simulations with the final model (best performer in the TL+AL+EKD model ensemble), compared with experiments\cite{gs1975density}, and simulation time lengths at a series of temperatures.}
  \begin{tabular}{lllll}
    % \toprule
    \hline
    T (K) & Time (ns) & Simulated density (g/ml) & Exp. (g/ml) \\
    % \midrule
    \hline
    248	& 3.5 & 0.993\(\pm\)0.001 &	0.990 \\
    268	& 3.0 & 1.002\(\pm\)0.001 & 0.999 \\
    277	& 3.0 & 1.002\(\pm\)0.001 & 1.000 \\
    298	& 2.0 & 0.998\(\pm\)0.001 & 0.997 \\
    318	& 1.0 & 0.987\(\pm\)0.002 & 0.990 \\
    338	& 2.0 & 0.976\(\pm\)0.001 & 0.981 \\

    % \bottomrule
    \hline
  \end{tabular}
  \label{tab:three-line}
\end{table}

\bibliography{references}

% Copy and paste your Supplemental Materials text here \cite{S_RefA}, blah, blah, blah, blah, blah, blah, ...
% \begin{equation}
%   i\hbar\frac{\partial}{\partial t}\psi(x,t) = -\frac{\hbar^2}{2m}\frac{\partial^2}{\partial x^2}\psi(x,t) + V(x,t) \psi(x,t)
% \end{equation}

% \begin{thebibliography}{11}
% \bibitem{S_RefA} A. Someone, C. Someone, D. Someone, Phys. Rev. Lett. {\bf 11}, 1111 (1911).
% \end{thebibliography}